\documentclass[twocolumn,times]{aastex63}
\usepackage{multirow}
\usepackage{subfigure}
\usepackage{booktabs}
\begin{document}

\title{Model constraints based on the IceCube neutrino non-detection of GRB 221009A}

\correspondingauthor{He Gao}
\email{gaohe@bnu.edu.cn}

\author[0000-0002-9165-8312]{Shunke Ai}
\affil{School of Physics and Technology, Wuhan University, Wuhan 430072, China}

\author[0000-0002-3100-6558]{He Gao}
\affiliation{Institute for Frontier in Astronomy and Astrophysics, Beijing Normal University, Beijing 102206, China;}
\affiliation{Department of Astronomy ,
Beijing Normal University, Beijing 100875, People's Republic of China;}

\begin{abstract}
GRB 221009A is a bright Gamma-ray burst (GRB) with isotropic energy being larger than $10^{54} ~{\rm ergs}$. Its fairly low redshift makes it a promising candidate for high energy neutrino detection. However, a neutrino search for this GRB reported by the IceCube collaboration yielded a null result. In this paper, we utilize the upper limit from IceCube observation to test different GRB prompt emission models. We find that, at least for this specific burst, the dissipative photosphere model could be ruled out in a large parameter space. The internal shock model can survive only with a large bulk motion Lorentz factor $\Gamma$, where the most stringent and conservative constraints are $\Gamma > \sim 450$ and $\Gamma > \sim 200$, respectively. Also, the ratio of the total dissipated energy that goes into the protons and electrons ($\epsilon_p / \epsilon_e$) can be constrained with a given $\Gamma$. For $\Gamma < 400$, $\epsilon_p / \epsilon_e < 10$ is required. For the Internal-collision-induced Magnetic Reconnection and Turbulence (ICMART) model, the constraint from GRB 221009A is modest. Under ICMART model, only for extreme situations when most dissipated energy deposit into protons and all accelerated protons are suitable for producing neutrinos, a slightly large bulk motion ($\Gamma > \sim 250$) is required. 
\end{abstract}


\section{Introduction}

After decades of investigations, the origin of GRB prompt emission is still under debate. For instance, the location of the prompt emission (i.e., the distance of GRB prompt emission from the central engine, $R$) is still not settled. Models invoking different jet composition have different emission radii \citep{zhang2018book}: for a matter-dominated scenario, the emission site would be the photosphere radius ($R_{\rm ph}\sim10^{11-12}$ cm; \cite{meszaros2000,peer2007}) or the internal shock radius ($R_{\rm IS}\sim10^{12-13}$ cm;  \cite{rees1994,daigne1998}); for a magnetically dominated scenario, the prompt emission would be powered at large radii (e.g., $\sim10^{15}$ cm) where significant magnetic dissipation occurs \citep{zhang2011,mcKinney2012,lazarian2019}. 

In the prompt emission sites where electrons are accelerated, protons are also expected to be accelerated, so that significant neutrino emission is possible from high-energy protons interacting with other baryons or with photons \citep{waxman1997,razzaque2003,dermer2003,guetta2004,murase2006a,murase2006b,hummer2012}. A guaranteed target photon source for the p$\gamma$ interaction is the GRB prompt emission itself, as long as the protons can be accelerated to an enough energy $E_p$ to satisfy the $\Delta$-resonance condition. Since high energy neutrinos suffer little from absorption effect along the propagation path, they provide valuable clues about the prompt emission sites, thus help to distinguish different GRB models \citep{zhang2013,pitik2021}. 

Over the years, the null results in the search for high energy neutrino signals coincident with GRBs in time and direction from the IceCube Collaboration, have made progressively non-detection upper limits on the neutrino flux from GRBs \citep{abbasi2010,abbasi2011,aartsen2015,aartsen2016,abbasi2022}, which places stringent constraints on the parameter space of GRB models \citep{li2012,he2012,gao2012,zhang2013}. Nevertheless, it has been shown that in some special individual cases, such as the nearby monster GRB 130427A, the non-detection of high energy neutrino could make even tighter constraints on the internal shock model and the photosphere model of GRBs \citep{gao2013}. 

Most recently, an extremely bright GRB, GRB 221009A, has attracted extensive attention. The burst first triggered the Fermi/Gamma-ray Burst Monitor (GBM) at 13:16:59 UT on 2022 October 9, with a fluence of $(2.12\pm0.05)\times10^{-5}{\rm erg~cm^{-2}}$ in 10–1000 keV within a duration of $T_{90}=327$ s \citep{GCN32642}. The object was then registered by the Swift Burst Alert Telescope (BAT) on 2022 October 9 at 14:10:17 UT, which is initiallly considered to be a Galactic X-ray transient (Swift J1913.1+1946 \citep{swiftGCN}). Its exceedingly bright Gamma-ray and X-ray flux enabled multiple ground- and space-based follow-up observations, allowing for extensive broadband afterglow monitoring from radio to very high energy $\gamma$-ray \citep{HXMT,GCN32638,GCN32656,GCN32660,GCN32677,GCN32694,GCN32658,GCN32676,GCN32736,GCN32750,GCN32757}, as well as rapid accurate determination of the event location and distance at redshift $z=0.151$, implying an isotropic radiation energy of $\sim2\times10^{54}~\rm erg$ \citep{GCN32648,GCN32686}. 

Comparing with GRB 130427A, GRB 221009A has even larger radiation energy and closer distance, making it a more promising candidate for neutrino detection. However, a neutrino search for this GRB reported by the IceCube collaboration still yielded a null result. Here we show that this null detection could provide interesting information about the properties of this GRB, and could make stringent constraints on the parameter space of various GRB models, including the dissipative photosphere model, internal shock model, as well as one representative magnetic dissipation model, i.e. the Internal-collision-induced Magnetic Reconnection and Turbulence (ICMART) model \citep{zhang2011}. 

\section{General formula for neutrino emission}

High energy neutrinos can be produced through the hadronic process when the accelerated protons interacts with other baryons or photons. We follow the procedure in \cite{zhang2013} to derive the formula for neutrino emission. 

Strong photo-meson interactions happen at the $\Delta$-resonance, when the proton energy $E_p$ and photon energy $E_\gamma$ (in the observer's frame) satisfy the threshold condition 
\begin{eqnarray}
E_p E_{\gamma} \gtrsim 0.160~{\rm GeV^2}~\left(\frac{\Gamma}{1+z}\right)^2,
\label{eq:delta_condition}
\end{eqnarray}
where $\Gamma$ represents the bulk motion Lorentz factor. There are several channels for $p \gamma$ interactions, of which $p\gamma\rightarrow\Delta^+\rightarrow n\pi^+$ and $p\gamma\rightarrow\Delta^+ \rightarrow p\pi^0$ are dominant with nearly equal cross-section \citep{kelner2008}. $\pi^+$ can subsequently decay to produce one positron and three neutrinos ($\nu_e$, $\nu_{\mu}$ and $\bar{\nu}_{\mu}$) with a typical neutrino energy $E_\nu \simeq 0.05 E_p$. For GRB photons with peak energy $E_{\gamma} \sim  1~{\rm MeV}$ in a radiation area with bulk motion Lorentz factor $\Gamma \sim 300$, the required photon energy for the neutrino production is $E_p \gtrsim 10~{\rm PeV}$, so that the corresponding neutrino energy should be $E_{\nu} \gtrsim 500~ {\rm TeV}$, which falls well into the detectable energy range of IceCube.

The photon number spectrum of GRB prompt emission can usually be fitted with a broken power law known as the Band function \citep{band1993}
\begin{eqnarray}
n_{\gamma}(E_\gamma) &=& \frac{dN_{\gamma}(E_{\gamma})}{E_{\gamma}} \nonumber \\
&=& n_{\gamma,b}  \left\{\begin{array}{cc}
\epsilon_{\gamma,b}^{\alpha} E_{\gamma}^{-\alpha}, & E_\gamma < \epsilon_{\gamma,b}\\
\epsilon_{\gamma,b}^{\beta} E_{\gamma}^{-\beta}, & E_\gamma \geq \epsilon_{\gamma,b},
\end{array}
\right.
\label{eq:photon_spectrum}
\end{eqnarray}
where $n_{\gamma,b}$ represents the specific photon number at $E = \epsilon_{\gamma,b}$. Assuming that the accelerated protons follow a power-law energy distribution $dN_p/dE_p \propto E_p^{-p}$, the neutrino spectrum produced through $p\gamma$ interaction can thus be estimated as \citep{waxman1997,abbasi2010}
\begin{eqnarray}
n_{\nu} (E_{\nu}) &=& \frac{dN_\nu(E_{\nu})}{dE_{\nu}} \nonumber \\
&=& n_{\nu,1} \left\{ \begin{array}{cc}
\epsilon_{\nu,1}^{\alpha_{\nu}} E_{\nu}^{-\alpha_\nu}, & E_{\nu} < \epsilon_{\nu,1}, \\
\epsilon_{\nu,1}^{\beta_\nu} E_{\nu}^{-\beta_\nu}, & \epsilon_{\nu,1} \leq E_{\nu} < \epsilon_{\nu,2},\\
\epsilon_{\nu,1}^{\beta_\nu} \epsilon_{\nu,2}^{\gamma_{\nu} - \beta_\nu} E_{\nu}^{-\gamma_\nu}, & E_{\nu} \geq \epsilon_{\nu,2},
\end{array}
\right.
\label{eq:neutrino_spectrum}
\end{eqnarray}
where the two breaks $\epsilon_{\nu,1}$ and $\epsilon_{\nu,2}$ are related to the break of Gamma-ray spectrum $\epsilon_{\gamma,b}$ and the synchrotron cooling of $\pi^{+}$ \footnote{The cooling of $\mu^+$ might be similar as $\pi^+$. However, about 1/3 of the total neutrinos expected from the p$\gamma$ interaction are produced directly by the decay of $\pi^+$, together with the generation of $\mu^+$. Therefore, although the cooling of $\mu^+$ might change the neutrino spectrum, its effects should not as significant as that from the cooling of $\pi^+$, so that we do not consider the cooling of $\mu^+$ in this paper.}, respectively, which would be detailed discussed later. $n_{\nu,1}$ represents the specific neutrino number at $E_{\nu} = \epsilon_{\nu,1}$. The power-law indices of the neutrino spectrum are related with the indices of photon spectrum as 
\begin{eqnarray}
\alpha_{\nu} = p + 1 - \beta,~\beta_\nu = p + 1 - \alpha,~\gamma_\nu = \beta_\nu + 2.
\end{eqnarray}
The expressions of $\alpha_{\nu}$ and $\beta_{\nu}$ are derived by assuming the neutrino spectrum is directly proportional to the distribution of accelerated protons and the $p\gamma$ optical depth (see Equation \ref{eq:tau}). The expression of $\gamma_{\nu}$ is obtained by solving the continuity equation for charged particles with synchrotron cooling, where the power-law index should plus 2 beyond the characteristic cooling energy \citep[e.g.][Chapter 5]{zhang2018book}.

The first break energy in the spectrum $\epsilon_{\nu,1}$ is corresponding to the p$\gamma$ interaction involving $E_{\gamma} = \epsilon_{\gamma,b}$, which can be calculated as
\begin{eqnarray}
\epsilon_{\nu,1} = 6.33 \times 10^5 ~{\rm GeV} \left(\frac{\epsilon_{\gamma,b}}{\rm MeV}\right)^{-1} \left(\frac{\Gamma}{300}\right)^2(1+z)^{-2}.&&\nonumber \\
&&
\end{eqnarray}
This expression is only valid when the portion of proton energy going into pion production (the pion production efficiency , denoted as $f_{\pi}$) is much smaller than $1$. Otherwise it should be modified through invoking the $p \gamma$ optical depth, which can be calculated as
\begin{eqnarray}
\tau_{p \gamma}^p = 8.9L_{{\rm GRB},52} \left(\frac{\Gamma}{300}\right)^{-2} R_{13}^{-1} \left(\frac{\epsilon_{\gamma,b}}{\rm MeV}\right)^{-1},
\end{eqnarray}
where $L_{\rm GRB}$ is the isotropic luminosity of the GRB. The convention $Q_x = Q/10^x$ is used and hereafter, where $Q$ can be any quantity. In this paper, we do not introduce the modified expression for $\epsilon_{\nu,1}$ but directly change the shape of the neutrino spectrum when $f_{\pi} \sim 1$, in which case the neutrino energy flux would no longer increase with higher $\tau_{p \gamma}$. In principle, the pion production efficiency should be estimated by comparing the pion production time scale ($t_{\pi}$) and the dynamical timescale ($t_{\rm dyn}$). Specifically, one should have  $f_\pi = {\rm min}\{1,t_\pi^{-1}/t_{\rm dyn}^{-1}\}$, where $t_{\rm dyn} \sim R/(\Gamma c)$ and \citep{stecher1968,murase2007}
\begin{eqnarray}
t_{\pi}^{-1} = \frac{c}{2\gamma_p^2} \int_{E_{\rm \gamma,th}}^{\infty} dE_{\gamma}^{\prime} \sigma_{p \gamma} \kappa E_{\gamma}^{\prime} \int_{E_{\gamma}^{\prime}/2\gamma_p}^{\infty} E_{\gamma}^{-2} n(E_{\gamma}) dE_{\gamma}.
\label{eq:t_pion}
\end{eqnarray}
$E_{\gamma}^{\prime}$ stands for the photon energy in the jet's rest frame. $n(E_{\gamma})$ is the specific photon number density in the jet's rest frame, which follows the power-law distribution in Equation \ref{eq:photon_spectrum}. $\gamma_p$ is the random motion Lorentz factor of photons. We find that $\tau_{p \gamma} \propto E_{\nu}^{m-1}$, where $m = \beta$ for $E_{\nu} < \epsilon_{\nu,1}$ and $m = \alpha$ for $E_{\nu} \geq \epsilon_{\nu,1}$. To be concise, we can write the pion production efficiency directly as a function of $\tau_{p \gamma}$, which reads as
\begin{eqnarray}
f_\pi = 1 - (1 - <\chi_{p \rightarrow \gamma}>)^{\tau_{p \gamma}}.
\label{eq:f_pi}
\end{eqnarray}
where $<\chi_{p \rightarrow \gamma}>  \simeq 0.2$ represents the average fraction of energy transferred from protons to pions. The $E_{\nu}$-dependent $p \gamma$ optical depth can be expressed as 
\begin{eqnarray}
\tau_{p \gamma}(E_{\nu}) = \tau_{p \gamma}^p \left(\frac{E_{\nu}}{\epsilon_{\nu,1}}\right)^{m-1}.
\label{eq:tau}
\end{eqnarray}
We chose $\tau_{p \gamma} = 5$\footnote{Note that Equation \ref{eq:f_pi} is only valid when $\tau_{p\gamma} < 3$ \citep{zhang2013}.
Suppose the energy of protons is responsible for $N$ rounds of pion production. $(1 - <\chi_{p \rightarrow \gamma}>)^{\tau_{p\gamma}}$ represents the fraction of energy that has not been transferred into poins, with $N \sim \tau_{p\gamma}$. For a greater $\tau_{p\gamma}$, $N\sim \tau_{p\gamma}^2$ is expected. Roughly, when $\tau_{p\gamma} = 5$, $f_\pi \approx 1 - (1 - <\chi_{p \rightarrow \gamma}>)^{\tau^2_{p \gamma}} \approx 0.99$, which has been close enough to $f_\pi =1$.} as the critical saturated optical depth where $f_\pi \sim 1$ and find the corresponding typical neutrino energy $\epsilon_{\nu,s}$. This typical energy will replace $\epsilon_{\nu,1}$ as the first break on the neutrino spectrum when $\epsilon_{\nu,s} < \epsilon_{\nu,1}$.

The second break in the neutrino spectrum $\epsilon_{\nu,2}$ is due to the synchrotron cooling of $\pi^{+}$ where the $\pi^{+}$ might cool down before decaying. The $\pi^{+}$ decay timescale is
\begin{eqnarray}
t_{\pi^{+},{\rm dec}}^{\prime} = \gamma_{\pi^+} \tau_{\pi^{+}} = 2.8 \times 10^{-8}s~\gamma_{\pi^+}.
\end{eqnarray}
where $\gamma_{\pi^+}$ represents the Lorentz factor for the charged particle's random motion in the jet's comoving frame. For relativistic charged pions, the synchrotron cooling timescle can be calculated as
\begin{eqnarray}
t_{\pi^+,{\rm syn}}^{\prime} = \frac{6\pi m_{\pi^+} c}{\gamma_{\pi^+} \sigma_{T,\pi^+} B^{\prime 2}}
\end{eqnarray}
where $m_{\pi^+} = 0.15 m_p$ represents the rest mass of charged pions. The Thomson scattering cross section of $\pi^+$ can be derived from that of the electrons as $\sigma_{T,\pi^+} = (m_{e} / m_{\pi^+})^2 \sigma_{T,e}$. The superscript $\prime$ means the physical quantities are defined in the jet's comoving frame and hereafter. The random magnetic can be obtained from the GRB luminosity, which is expressed as
\begin{eqnarray}
B^{\prime} &=& \left[\frac{L_{\rm GRB} (\epsilon_B/\epsilon_e)}{2 R^2 \Gamma^2 c
}\right]^{1/2},
\end{eqnarray}
Here we assume a fraction $\epsilon_e$ of the dissipated energy goes into the electrons and is all radiated as $\gamma$-rays, a fraction $\epsilon_p$ of the dissipated energy goes into the protons, and a fraction $\epsilon_B$ of the dissipated energy goes into the random magnetic field. When $t_{\pi^+,{\rm dec}}$ equals to $t_{\pi^+,{\rm syn}}$, the corresponding $\gamma_{\pi^+,2}$ can be derived. Since the energy of $\pi^{+}$ would be shared nearly equally by four leptons, the second break on the neutrino spectrum is calculated as
\begin{eqnarray}
\epsilon_{\nu,2} &=& \frac{1}{4} {\cal D} \gamma_{\pi^+,2} m_{\pi^+} c^2 \nonumber \\
&=& 1.17\times 10^8 ~ {\rm GeV} L_{\rm GRB,52}^{-1/2}  \epsilon_{e,-1}^{1/2} \nonumber \\ 
&&R_{13} \left(\frac{\Gamma}{300}\right)^2 (1+z)^{-1} \epsilon_{B,-1}^{-1/2}.
\end{eqnarray}
where ${\cal D} \approx 2\Gamma$ is the Doppler factor.

The normalization parameter $n_{\nu,1}$ can be estimated from
\begin{eqnarray}
 \int_{\epsilon_{\nu,1}}^{\epsilon_{\nu,2}} n_{\nu} E_{\nu} dE_{\nu} = \frac{1}{8} \int_{E_{p,1}}^{E_{p,2}} f_\pi E_{p} \frac{dN_p}{dE_p}dE_p,
 \label{eq:normalization}
\end{eqnarray}
where the factor $1/8$ is the production of $1/2$ and $1/4$, because on average half of the protons would produce $\pi^+$ and the energy of $\pi^+$ would be shared nearly equally by four leptons. Also, one can derive
\begin{eqnarray}
\frac{dN_p}{dE_p} = \frac{(\epsilon_p/\epsilon_e)E_{\rm GRB}}{{\rm ln}(E_{p, {\rm max}}/E_{p, {\rm min}})} E_p^{-p},
\end{eqnarray}
for $p = 2$. $E_{p,1}$ represents the energy of protons which can most efficiently participate in the $\Delta^{+}$-resonance with photons with $E_{\gamma} = \epsilon_{\gamma,b}$ while $E_{p,2}$ is the proton energy related to the neutrino energy of $\epsilon_{\nu,2}$ \footnote{If the pion cooling is not significant, the second break on the neutrino spectrum should be determined by $E_{p,{\rm max}}$.}. $E_{\rm GRB}$ is the observed isotropic energy of the GRB.  $E_{p,{\rm max}}$ and $E_{p,{\rm min}}$ represent the maximum and minimum energy of accelerated protons, respectively, in the observer's frame. $E_{p, {\rm max}}$ could be estimated by equaling the dynamical timescale $t_{\rm dyn}^{\prime} \sim R / \Gamma c$ with the accelerating timescale of protons $t_{\rm acc}^{\prime} \sim E_p^{\prime} / (e B^{\prime} c)$.
Eventually, we have 
\begin{eqnarray}
E_{p, {\rm max}}  \leq 8.0 \times 10^{11}~{\rm GeV} (\epsilon_B/\epsilon_e)^{1/2} L_{{\rm GRB},52}^{1/2} \left(\frac{\Gamma}{300}\right)^{-1}.&& \nonumber \\
&&
\end{eqnarray}
A lower limit for minimum proton energy can be set as $E_{p, {\rm min}} > \Gamma m_p c^2 = 2.7 \times 10^{2} ~{\rm GeV} ~(\Gamma/300) $.
Here we define
\begin{eqnarray}
f_p = \frac{\int_{E_{p,1}}^{E_{p,2}} E_p \frac{dN_p}{dE_p}dE_p}{\int_{E_{p,{\rm max}}}^{E_{p,{\rm max}}} E_p \frac{dN_p}{dE_p}dE_p}.
\end{eqnarray}
The physical meaning of $f_p$ is the fraction of accelerated protons that have energy ranging from $E_{p,1}$ to $E_{p,2}$, which can most efficiently participate in the $p \gamma$ interaction.
For $p = 2$, the expression of $f_p$ can be derived as
\begin{eqnarray}
f_p &=& \frac{{\rm ln}(E_{p,1}/E_{p,2})}{{\rm ln}(E_{p, {\rm max}}/E_{p, {\rm min}})} \nonumber \\
&=& \frac{{\rm ln}(\epsilon_{\nu, 1}/\epsilon_{\nu,2})}{{\rm ln}(E_{p, {\rm max}}/E_{p, {\rm min}})}.
\end{eqnarray}
%where $E_{p,2}$ is the proton energy related to the neutrino energy of $\epsilon_{\nu,2}$ \footnote{If the pion cooling is not significant, the second break on the neutrino spectrum should be determined by $E_{p, {\rm max}}$.}. 
With the constraints on $E_{p, {\rm max}}$ and $E_{p, {\rm min}}$, we have $f_p > \sim 0.2$. 

Then, the right-hand side of Equation \ref{eq:normalization} can be rewritten as
\begin{eqnarray}
\frac{1}{8} \int_{E_{p,1}}^{E_{p,2}} f_\pi E_{p} \frac{dN_p}{dE_p}dE_p &\approx& \frac{1}{8} f_\pi \int_{E_{p,1}}^{E_{p,2}} E_{p} \frac{dN_p}{dE_p}dE_p \nonumber \\
&=& \frac{1}{8} f_{\pi} f_p (\epsilon_p/\epsilon_e) E_{\rm GRB}.
\end{eqnarray}
Here $f_{\pi}$ is approximately treated as constant during the integration, because when caring about neutrinos with energy beyond $\epsilon_{\nu,1}$, $\tau_{p\gamma} \approx \tau_{p\gamma}^p$ for $m = \alpha \approx 1$ (see Equation \ref{eq:tau}). Assuming $\alpha = 1$, the left-hand side of Equation \ref{eq:normalization} can be rewritten as
\begin{eqnarray}
 \int_{\epsilon_{\nu,1}}^{\epsilon_{\nu,2}} n_{\nu} E_{\nu} dE_{\nu} =  E_{\nu,1}^2 n_{\nu,1} {\rm ln}(\epsilon_{\nu,2}/\epsilon_{\nu,1})
\end{eqnarray}
The specific neutrino number at $E_{\nu} = \epsilon_{\nu,1}$ can thus be calculated as
\begin{eqnarray}
n_{\nu, 1} &\approx&  \frac{1}{8} \frac{f_\pi f_p(\epsilon_p/\epsilon_e)E_{\rm GRB}}{{\rm ln}(\epsilon_{\nu,2}/\epsilon_{\nu,1})} \epsilon_{\nu,1}^{-2},
\end{eqnarray}
for $p = 2$ and $\alpha = 1$. When $\epsilon_{\nu,s} < \epsilon_{\nu,1}$, the spectrum is modified as
\begin{eqnarray}
n_{\nu} (E_{\nu}) =
n_{\nu,s} \left\{ \begin{array}{cc}
\epsilon_{\nu,s}^{\alpha_{\nu}} E_{\nu}^{-\alpha_\nu}, & E_{\nu} < \epsilon_{\nu,s}, \\
\epsilon_{\nu,s}^2 E_{\nu}^{-2}, & \epsilon_{\nu,s} \leq E_{\nu} < \epsilon_{\nu,2},\\
\epsilon_{\nu,s}^2 \epsilon_{\nu,2}^{\gamma_{\nu} - 2} E^{-\gamma_\nu}, & E_{\nu} \geq \epsilon_{\nu,2},
\end{array}
\right.
\end{eqnarray}
where the indexes $\alpha_{\nu}$ and $\gamma_{\nu}$ are adopted as the same as that used in Equation \ref{eq:neutrino_spectrum}. $n_{\nu,s}$ represents the neutrino specific number at $E_{\nu} = \epsilon_{\nu,s}$, which can be calculated as
\begin{eqnarray}
n_{\nu,s} &\approx& \frac{1}{8}  \frac{ f_p(\epsilon_p/\epsilon_e)E_{\rm GRB}}{{\rm ln}(\epsilon_{\nu,2}/\epsilon_{\nu,s})} \epsilon_{\nu,s}^{-2}.
\end{eqnarray}
Overall, the observed neutrino fluence can be calculated as
\begin{eqnarray}
\phi_{\nu} (E_{\nu}) = \frac{E_{\nu}^2 n_{\nu}(E_{\nu})}{4\pi D^2},
\end{eqnarray}
where $D$ is the distance from the GRB source to the observer.

\section{Model Constraints}
Since GRB 221009A is such a bright and close event, the non-detection of associated neutrinos suggests that the location of the prompt emission should be relatively large, in order to make the number density of $\gamma$-ray photons in the radiation area not high enough to participate in the $p \gamma$ interaction to produce high energy neutrinos. The prompt emission site is highly model dependent:
\begin{itemize}
    \item I. Dissipative photosphere model: Here we consider the photosphere radius related to the Thompson scattering for the $\gamma$-ray photons, which is estimated as $R \approx 3.0\times 10^{13} L_{{\rm GRB},52} \epsilon_{e,-1}^{-1} (\Gamma/300)^{-3}$. 
    \item II. Internal shock model: The typical radius for the $\gamma$-ray radiation is roughly estimated as $R \sim 2.7 \times 10^{14}~{\rm cm}~ (\Gamma/300)^2 (\delta t_{\rm min} / 0.1~{\rm s}) (1+z)^{-1}$ where $\delta t_{\rm min}$ represents the minimum variability time scale in the GRB observation. 
    \item III. ICMART model: The typical radius of the ICMART model is roughly estimated as $R \sim 2.7 \times 10^{15}~{\rm cm}~ (\Gamma/300)^2 (\delta t_{\rm slow} / 1~{\rm s}) (1+z)^{-1}$ where $\delta t_{\rm slow} \gtrsim 1~{\rm s}$ stands for the slow-variation timescale for the GRB lightcurves.
\end{itemize}

\begin{figure}
\resizebox{75mm}{!}{\includegraphics[]{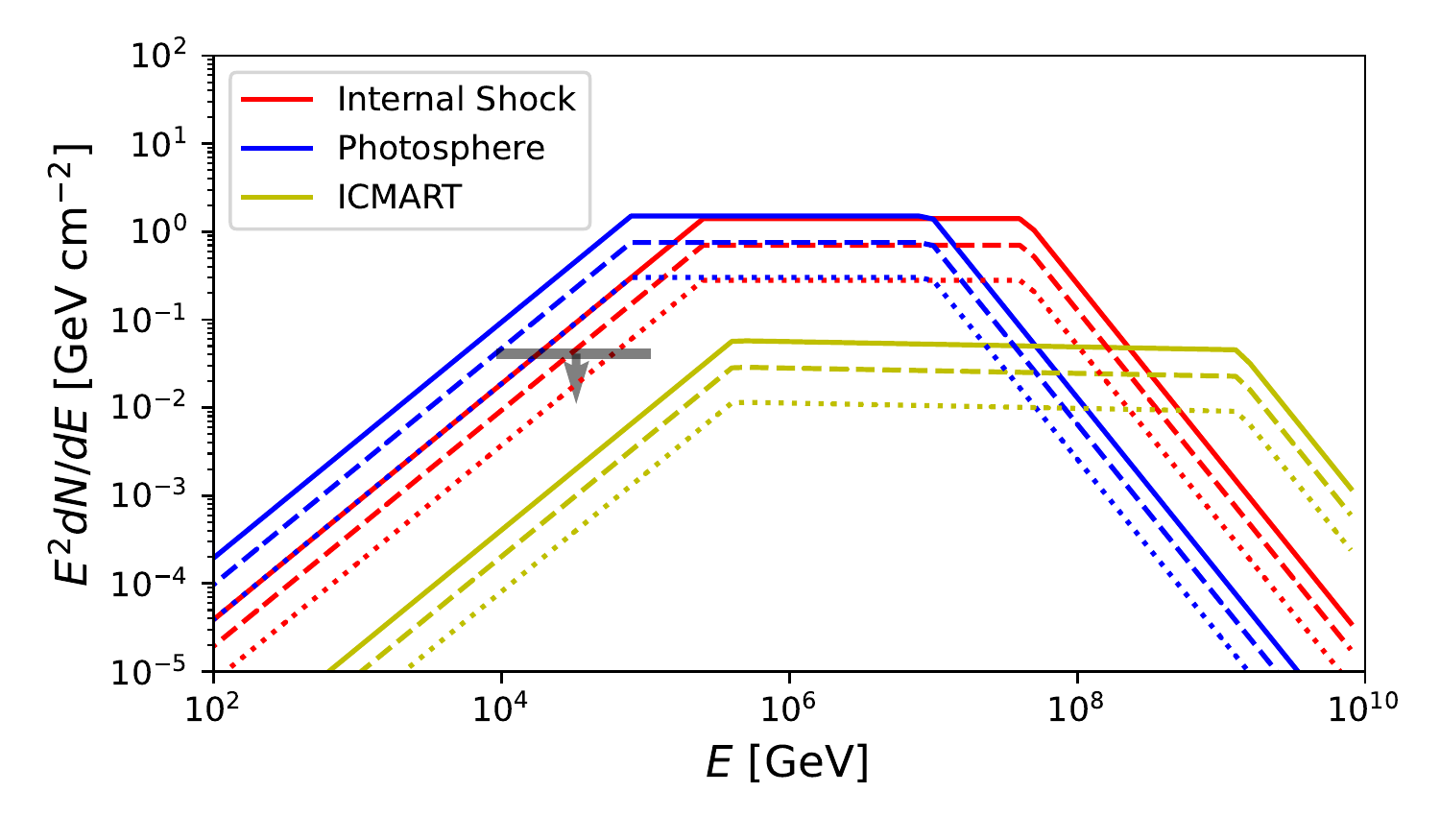}} \\
\resizebox{75mm}{!}{\includegraphics[]{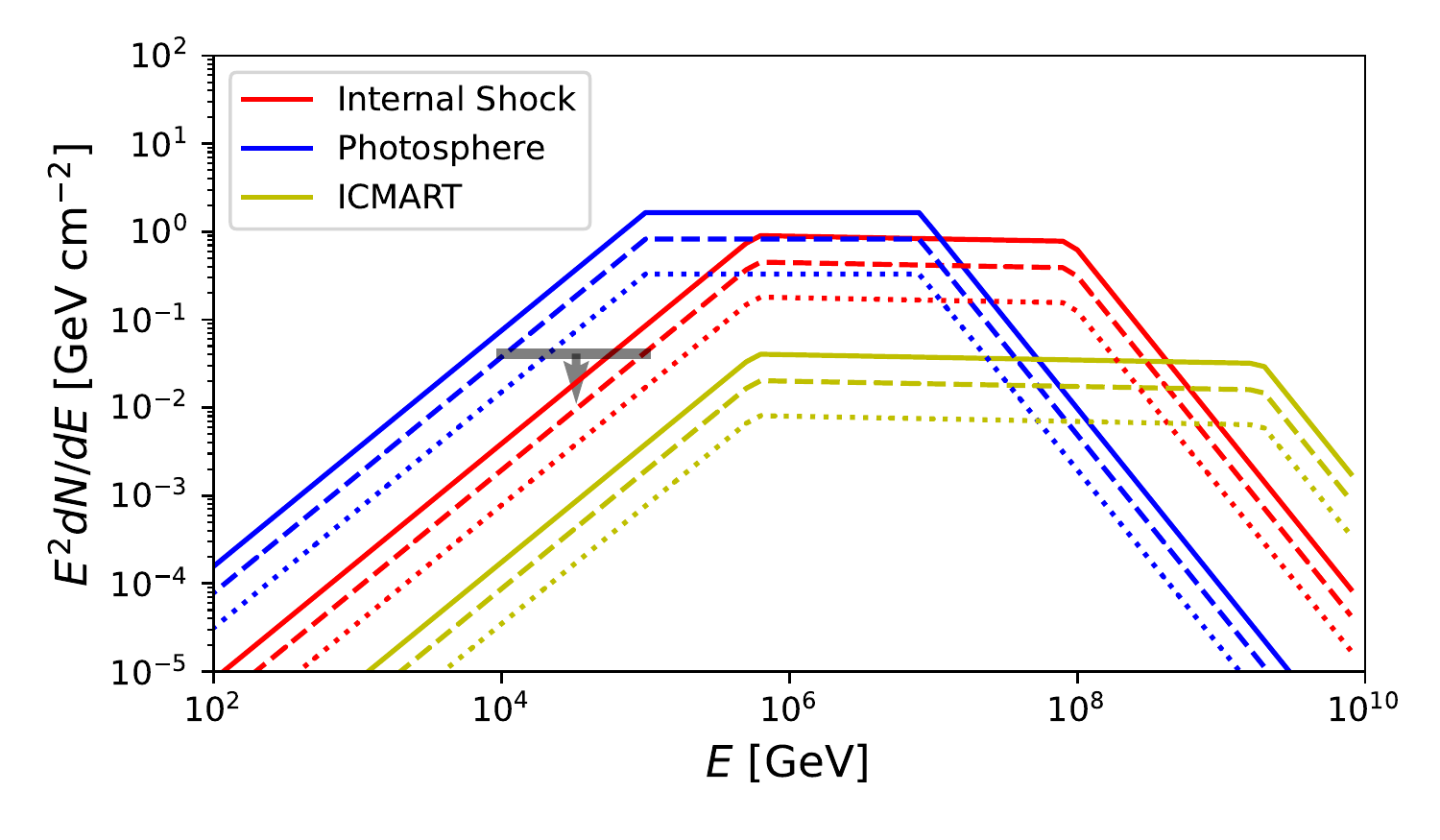}} \\
\resizebox{75mm}{!}{\includegraphics[]{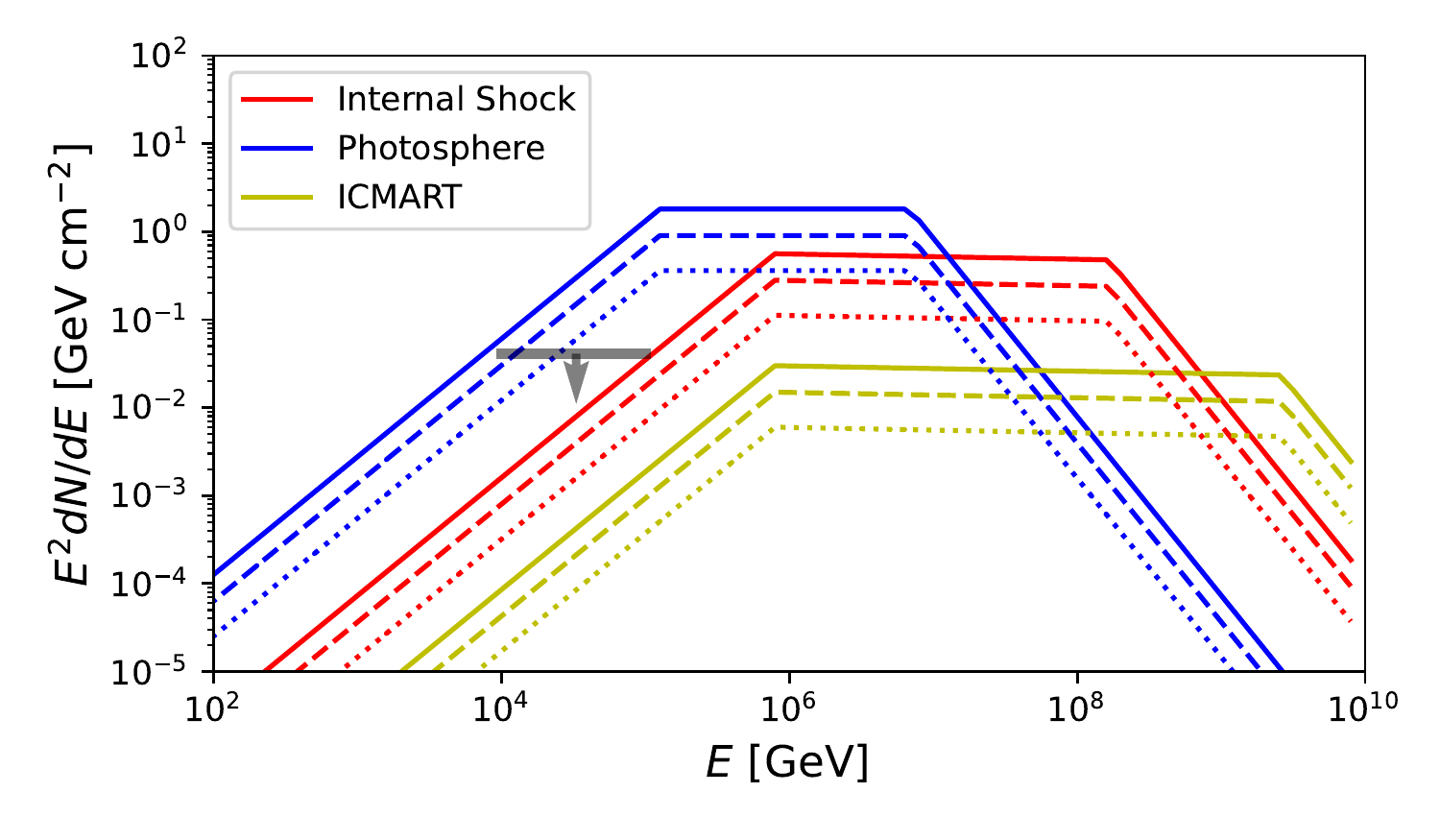}} \\ 
\resizebox{75mm}{!}{\includegraphics[]{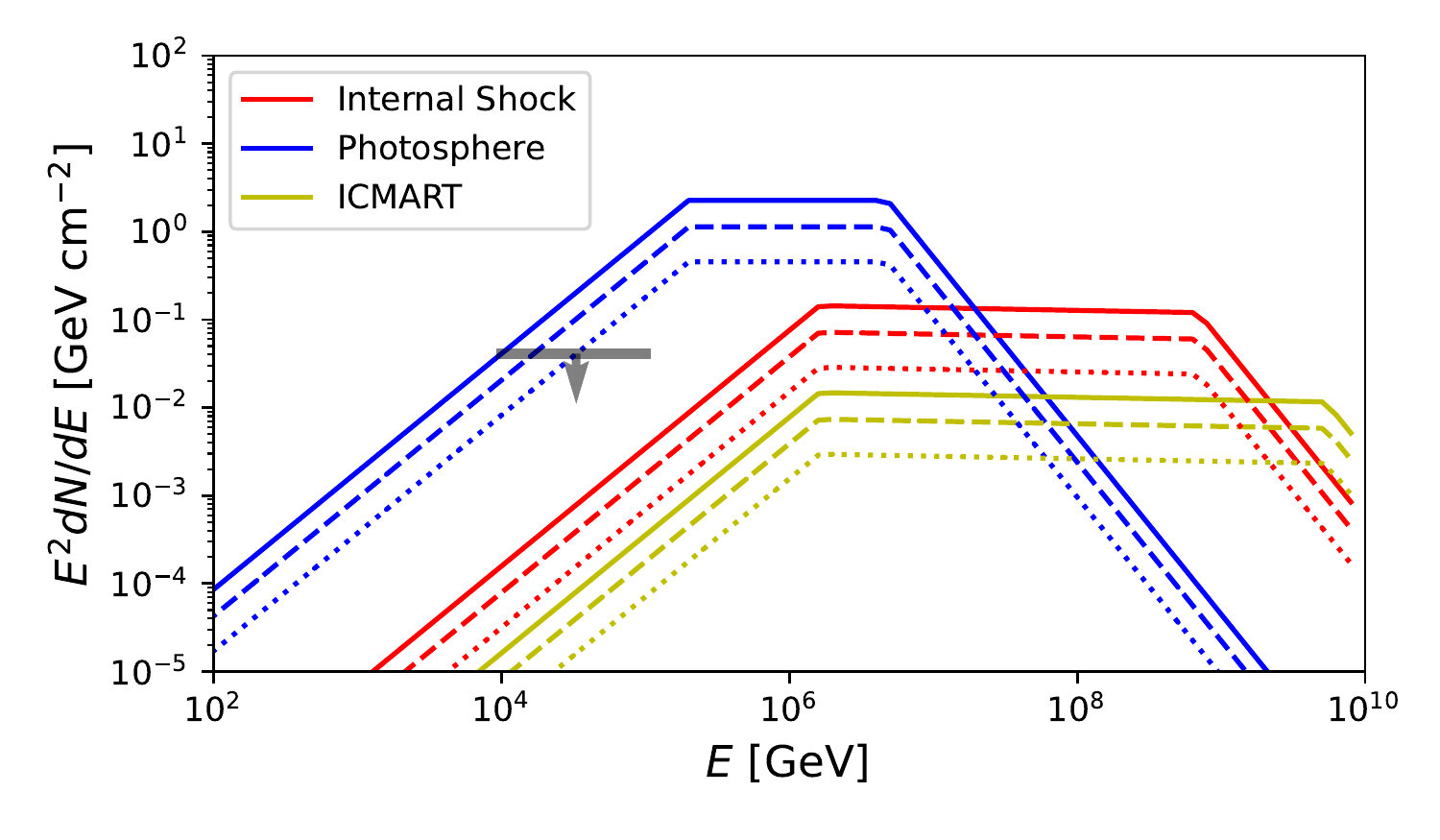}} \\ 
\caption{The predicted neutrino spectrum for the internal shock, photosphere and ICMART model. From the upper to the lower panel, $\Gamma = 250$, $300$, $350$ and $500$ are adopted, respectively. The solid, dashed and dotted lines stand for $f_p = 1.0$, $0.5$ and $0.3$, respectively. For GRB 221009A, the indexes for the Band function are fitted as $\alpha = 0.97$ and $\beta = 2.34$ and the isotropic luminosity is estimated as $L_{\rm GRB} \sim 1.9 \times 10^{52}~{\rm erg~s^{-1}}$. For the internal shock model, $\delta t_{\rm min} = 0.01~{\rm s}$ is assumed. For the ICMART model, $t_{\rm slow} = 1~{\rm s}$ is assumed. For all the panels, $\epsilon_B/\epsilon_e = 1.0$ and $\epsilon_p/\epsilon_e = 3.0$ are adopted. The upper limit of neutrino fluence set by IceCube observation is shown as the grey vertical line.}
\label{fig:neutrino_flux}
\end{figure}
Based on the model described in Section 2, the predicted neutrino fluxes associated with GRB221009A from the internal shock, dissipative photosphere and ICMART models are shown in Figure \ref{fig:neutrino_flux}.  It is clear that the upper limit of the neutrino fluence given by IceCube observation, $E_{\nu}^2 \phi_{\nu} < 4.1 \times 10^{-2}~{\rm GeV~cm^{-2}}$ , can put constraints on the parameters for different GRB prompt emission models. Consider that IceCube is most sensitive to detect neutrinos with energy from $10$ TeV to $100$ TeV. Therefore, the upper limit for the neutrino fluence introduced above is only valid in this energy range, outside which the constraints should be highly relaxed.

We first test some fiducial situations: we adopt a typical value for the bulk Lorenz factor $250\leq\Gamma\leq500$, and set $\epsilon_p/\epsilon_e=3$. Considering that $f_p$ and $\epsilon_p/\epsilon_e$ are with large uncertain and difficult to be determined from the observations, here we test various situations with $0.2\leq f_p \leq1$ and constraints $\epsilon_p/\epsilon_e$ individually with the internal shock model assumed later. We find that
\begin{itemize}
    \item the dissipative photosphere model is highly disfavored, since its predicted neutrino fluence always exceeds the IceCube limit even for a relatively small $f_p$ ($f_p = 0.2$). The result still stands for a lower $\epsilon_p/\epsilon_e$ value (e.g. $\epsilon_p/\epsilon_e < 1$).
    \item the internal shock model is restricted to a certain extent. When $\Gamma \lesssim 250$, the neutrino fluence predicted by the internal shock model will always violate the non-detection result from IceCube. When $\Gamma = 300$, the internal shock model could survive only with a relatively low $f_p$ ($f_p<0.5$). When $\Gamma \gtrsim 350$, the predicted neutrino fluence will be reduced below the IceCube limit even for a large $f_p$ ($f_p = 1$). The survival parameter space would become wider for a lower $\epsilon_p/\epsilon_e$ value. 
    \item the ICMART model always predicts neutrino fluence lower than the upper limit from IceCube. 
\end{itemize}

\begin{figure}
\resizebox{75mm}{!}{\includegraphics[]{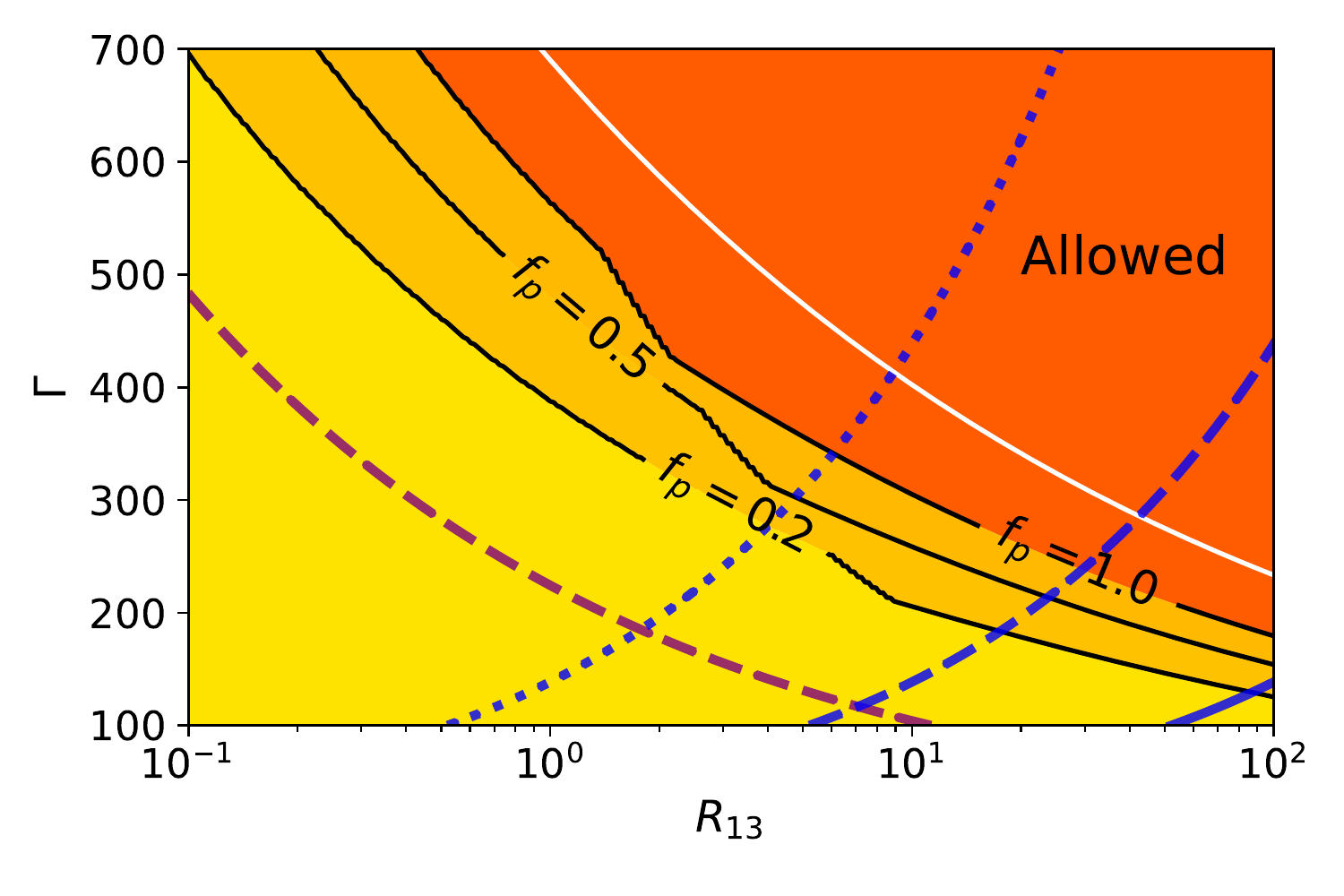}} \\
\resizebox{75mm}{!}{\includegraphics[]{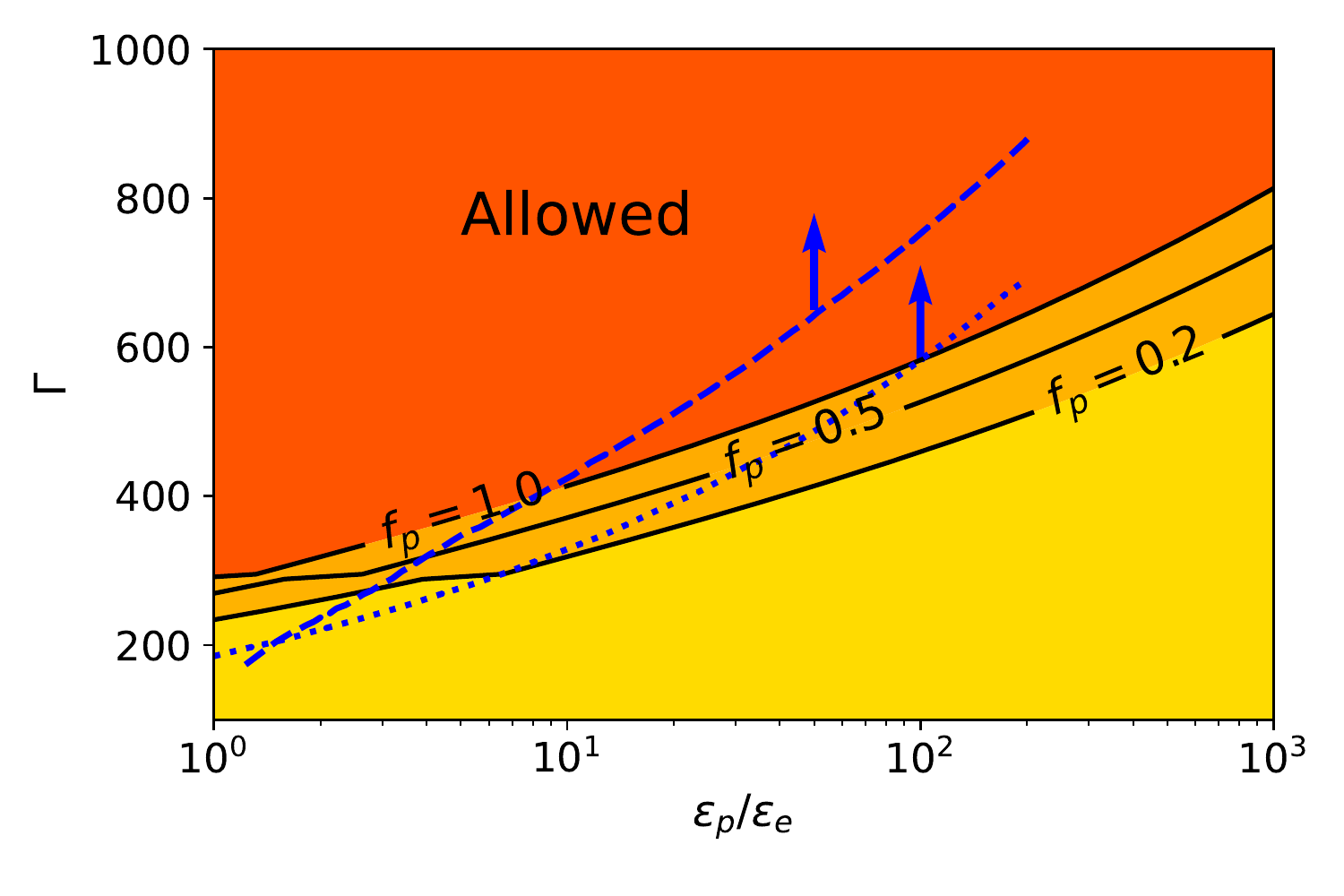}}
\caption{Upper panel: The allowed and excluded regions on the $R$-$\Gamma$ plot. $\epsilon_p/\epsilon_e = 3$ is adopted. The cases with different $f_p$ are shown with different black lines. The red dashed lines show the $R$-$\Gamma$ dependence in the dissipative  photosphere model. The blue solid, dashed and dotted lines represent the internal shock model with $\delta t_{\rm min} = 1s$, $0.1s$ and $0.01s$ assumed, respectively. As a comparison, the white solid line shows the constraint when $\epsilon_p / \epsilon = 10$ and $f_p = 1$ are adopted. Lower panel: The allowed and excluded regions on the ($\epsilon_p/\epsilon_e$)-$\Gamma$ plot in the internal shock scenario with $\delta t_{\rm min} = 0.01s$. The constraints with $90\%$ C.L. obtained from the non-detection of IceCube's stacking searches for GRBs are also shown on this plot. The blue dashed and dotted lines are extracted from \cite{aartsen2017} and \cite{lucarelli2022}, respectively.}

\label{fig:contour}
\end{figure}
To be more general, we plot the model-independent allowed and excluded $R$-$\Gamma$ space in the upper panel of Figure \ref{fig:contour} with $\epsilon_p/\epsilon_e = 3$ adopted. Obviously, the region with lower $R$ and $\Gamma$ would be ruled out, which is reasonable because in this parameter space the $\gamma$-ray number density is more sufficient for the neutrino production. As the values of $f_p$ decreases, the restrictions on $R$ and $\Gamma$ become weaker. We also show the $R$-$\Gamma$ dependence for different models on the plot and find that:
\begin{itemize}
    \item given the luminosity of GRB 221009A, the line representing the dissipative photosphere model always locates in the exclusion region;
    \item the lines representing the internal shock model cross both the allowed and exclusive regions, indicating that the internal shock model could survive, but large values for $\Gamma$ and $R$ are required. The most stringent constraint could be placed when $f_p = 1.0$ and $\epsilon_p / \epsilon_e = 10$. In this case, $\Gamma$ is required to be larger than $\sim450$, when $\delta t_{\rm min}$ is in order of 0.01 s. Note that the minimum timescale for GRB 221009A is rather difficult to determine, since the burst is so bright that most gamma ray detectors are saturated. If the real $\delta t_{\rm min}$ is larger than 0.01 s, the allowed value of $\Gamma$ would become smaller.
    \item since the typical emission radius of the ICMART model is $\sim 10^{15} {\rm cm}$, the most stringent constraint could be placed to $\Gamma > \sim 250$ when $f_p = 1.0$ and $\epsilon_p/\epsilon_e = 10$. Almost no effective constraint could be placed for lower values of $f_p$ and $\epsilon_p/\epsilon_e$. 
\end{itemize}

With the internal shock model assumed, the constraints in the $(\epsilon_p/\epsilon_e)$-$\Gamma$ space are plotted in the lower panel of Figure \ref{fig:contour}. We find that, when $\delta t = 0.01~s$ is adopted, the constraints here is comparable with that from IceCube's stacking searches for GRBs \citep{aartsen2017,lucarelli2022}. When $f_p = 1$ is adopted, $\epsilon_p / \epsilon_e < \sim 10$ is required for $\Gamma = 400$; $\epsilon_p / \epsilon_e < \sim 3$ is required for $\Gamma = 350$; $\epsilon_p / \epsilon_e < \sim 1$ is required for $\Gamma = 300$.

\section{Conclusion and Discussion}
GRB 221009A is a bright long Gamma-ray burst that occurs in the near universe. The non-detection of IceCube neutrinos indicates that the neutrino production from GRB event is inefficient, so that some prompt emission models predicting high neutrino flux should be disfavored or even ruled out. Here we tested three prompt emission models for GRBs, including the internal shock model, the dissipative photosphere model and the ICMART model. 

With the upper limit of neutrino fluence given by the IceCube observation, we find that, at least for the specific case of GRB 221009A, the dissipative photosphere model would be ruled out within a large range of parameter space. The internal shock model could survive under limited parameter spaces, where a relatively large bulk motion Lorentz factor (e.g., $\Gamma > 200 \sim 450$) is required. The constraint on ICMART model is rather modest. For lower values of $f_p$ and $\epsilon_p/\epsilon_e$, almost no effective constraint could be placed. Only for extreme situations when $f_p \sim 1.0$ and $\epsilon_p/\epsilon_e \sim 10$ (i.e. most dissipated energy deposit into protons and all accelerated protons are suitable for producing neutrinos), a slightly large bulk motion ($\Gamma > \sim 250$) is required if the ICMART event happens at $R=10^{15}$ cm from the central engine. 

Overall, although the internal shock model and ICMART model cannot be completely distinguished, the IceCube neutrino non-detection indicates that GRB 221009A likely has a large bulk motion Lorentz factor, which is consistent with the expectations based on other observations, e.g., the onset timescale being earlier than the very first optical detection at $\sim 3000$ s. 

It is worth noting that, based on the IceCube's non-detection of track-like neutrinos, our constraints from GRB 221009A is slightly more stringent than the earlier constraints from GRB 130427A \citep{gao2013}, due to a closer distance. Our constraints are comparable with that from IceCube's stacking searches for GRBs \citep{aartsen2017,lucarelli2022}, as shown in the lower panel of Figure \ref{fig:contour}. Detailed studies based on the effective detecting area of IceCube in different energy band can make the constraints more stringent, especially when large $\Gamma$ is considered, with which the neutrino fluence peaks at a higher energy. In this case, the upper limit for the neutrino fluence in the $10$ TeV - $100$ TeV used in this paper would be less efficient. When relatively low $\Gamma$ value is considered, our constraints become more stringent than those from the stacking searches for GRBs. In addition, a recent work \citep{murase2022} also discussed the constraints on GRB models based on the non-detection of neutrinos from GRB 221009A through a different approach. They calculated the neutrino number flunece in each energy band and integrated them over the whole spectrum as well as the IceCube's effective detecting area to estimated neutrino numbers expected to be detected. Finally, their results are highly consistent with ours.

In addition, we should note that all the calculation and discussions above are based on the "one-zone" assumption, which means the protons are accelerated in the same zone as where the Gamma-ray photons are emitted. If multi-zone models are considered \citep{bustamante2015,heinze2020}, the constraints might be relaxed.

\section*{Acknowledgments}
We thank Bing Zhang and Yuanhong Qu for helpful discussion. We thank Walter Winter, Kohta Murase and the anonymous referee for useful comments. This work is supported by the National Natural Science Foundation of China (Projects:12021003). We acknowledge the science research grants from the China Manned Space Project with NO. CMS-CSST-2021-A13 and CMS-CSST-2021-B11.

\end{document}